\documentclass[12pt]{article}
\usepackage{times}
\usepackage{graphicx,amssymb,amsmath,epsfig,color,ulem}
\usepackage{tabularx}

\def\apjl{ApJL }

\def\apj{ApJ }

\def\apjs{ApJS }
\def\araa{ARAA }

\def\nat{Nature }
\def\mnras{MNRAS }
\def\prd{Phys. Rev. D. }

\def\azh{Astron. J.}

\newcounter{lastnote}

\title{Relativistic Jets in Core Collapse Supernovae}

\author
{Tsvi Piran,$^{1\ast}$ Ehud Nakar,$^{2}$, Paolo Mazzali,$^{3,4}$, Elena Pian$^{5,6}$\\
\\
\normalsize{$^{1}$Racah Institute of Physics, The Hebrew University of Jerusalem, Jerusalem 91904, Israel}\\
\normalsize{$^{2}$Raymond and Beverly Sackler School of Physics \&
Astronomy, Tel Aviv University, Tel Aviv 69978, Israel}\\
\normalsize{$^{3}$Astrophysics Research Institute, Liverpool John Moores University, IC2, Liverpool L3 5RF, UK} \\
\normalsize{$^{4}$Max-Planck-Institut fur Astrophysik, Karl-Schwarzschild-Str. 1, D-85748 Garching, Germany}\\
\normalsize{$^{5}$INAF, Istituto di Astrofisica Spaziale e Fisica Cosmica di Bologna, I-40129 Bologna, Italy} \\
\normalsize{$^{6}$Scuola Normale Superiore, Piazza dei Cavalieri 7, I-56126 Pisa, Italy}\\
\\
\normalsize{$^\ast$E-mail:  tsvi@phys.huji.ac.il}
}

\date{}

\begin{document} 


\baselineskip24pt


\maketitle


\begin{abstract}
After several decades of extensive research the mechanism driving core-collapse supernovae (CCSNe) is still unclear.  A common mechanism is a neutrino driven outflow, but others have been proposed.  Among those, a long-standing idea is that jets play an important role in SN explosions.  Gamma-ray bursts (GRBs) that accompany rare and powerful  CCSNe, sometimes called €˜``hypernovae", provide a clear evidence for a jet activity. The relativistic GRB jet  punches a hole in the stellar envelope and produces the observed gamma-rays far outside the progenitor star. While SNe and jets coexist in long GRBs, the relation between the mechanisms driving the hypernova and the jet is unknown. Also unclear is the relation between the rare hypernovae and the more common CCSNe.  Here we {present observational evidence that indicates} that choked jets are active in CCSNe types that are not associated with GRBs. A choked jet deposits all its energy in a cocoon. The cocoon eventually breaks out from the star releasing energetic material at very high, yet sub-relativistic, velocities.   This fast moving material has a unique signature that can be detected in early time  SN spectra.  We find a clear evidence for this signature in several  CCSNe,  all involving progenitors that have lost all, or most, of their hydrogen envelope prior to the explosion. These include CCSNe that don't harbor GRBs or any other relativistic outflows. Our findings suggest a continuum of central engine activities in different types of CCSNe and call for rethinking of the explosion mechanism of regular CCSNe.     
\end{abstract}


Massive stars end their lives in supernova (SN) explosions releasing typically $\sim 10^{51}$ ergs (sometimes called FOE or Bethe) in kinetic energy and a fraction of that in a visible light. 
 As the star consumes its energy reservoir its core collapses (hence Core Collapse Supernova - CCSNe) and becomes a compact object. A shock wave that propagates outwards ejects the envelope and synthesizes radioactive $^{56}$Ni that powers part of the visible SN light.
So far,  in addition to the explosions themselves, 
 we have seen the massive stellar progenitors, neutrinos produced by the newborn neutron star, the compact objects left behind (typically a neutron star) and the expanding matter, {(the supernova remnant)}.
 All these observations confirm the general picture outline by Baade and Zwicky already in the 1930's \cite{BaadeZwicky}.  However, while the basic picture  is well understood, in spite of several decades of research, the mechanism(s) powering the shocks that drive the SNe  is not clear.
Models suggested (see e.g. \cite{Janka12} and references therein) include neutrino heating, magnetohydrodynamic, thermonuclear, bounce-shock, acoustic and phase transition mechanisms. 
The neutrino driven explosion, possibly in combination with hydrodynamic non-spherical instabilities and non-radial flows,
is the current favorite (at least for most common core collapses,  type II SNe), while others (e.g., bounce-shock) seems highly unlikely.  In spite of the importance of 3D effects, the neutrino driven explosion is supposed to produce roughly spherical explosions. 
Among the other mechanisms  a long-standing idea, proposed already in the early 1970's \cite{LeBlancWIlson,Bisnovatyi,OstrikerGunn}, is that jets (particularly magnetically driven ones)  play an important role in SN explosions. Here we explore observational evidence of this idea. 

Rare and powerful (typically $10^{52}$ ergs) CCSNe, sometime called Hypernovae, accompany long Gamma-Ray Bursts (GRBs) (see e.g.\cite{Woosley06}).  These explosions involve  two distinct  components: a narrowly collimated relativistic jet that produces the GRB (see e.g. \cite{P04} and references therein) and a more isotropic (yet not necessarily spherically symmetric) massive SN explosion. The SN ejecta typically carries $\sim$10-100 times more energy than the GRB jet (see e.g. \cite{mazzali2014} and references therein).  Thus, while the jet itself cannot drive the SN explosion, it is reasonable to expect that the central rapidly rotating compact object that must be present at the center of the collapsing star to drive the GRB jet, is related to the energy source that 
drives the SN explosion. In GRBs the jet successfully penetrates the massive stellar envelope and we observe its emission directly. 

The association of SNe with GRBs bring up several important questions. First, are there hypernova where the GRB jets fail to breakout, namely choked within the stellar envelope. Second, do hidden jets exist in other types of CCSNe as well, and if they do can we detect them?  Finally,  what is the relation, if any, between the explosion mechanism of GRB associated SNe and other types of SNe. We address these questions here. We first establish a clear observational signature of hidden jets. This signature can be detected in the early (first few days) spectra of CCSNe, provided that those arise in stars that have lost all (or almost all) of their heavy hydrogen envelopes prior to the SN explosion, namely in type Ib/c, and possibly IIb, SNe.  We then proceed to demonstrate that this signature has already been observed in several SNe and that it enables us to estimate the jet parameters (its total energy and opening angle).

\begin{figure}[!ht]
\includegraphics[width=1\textwidth]{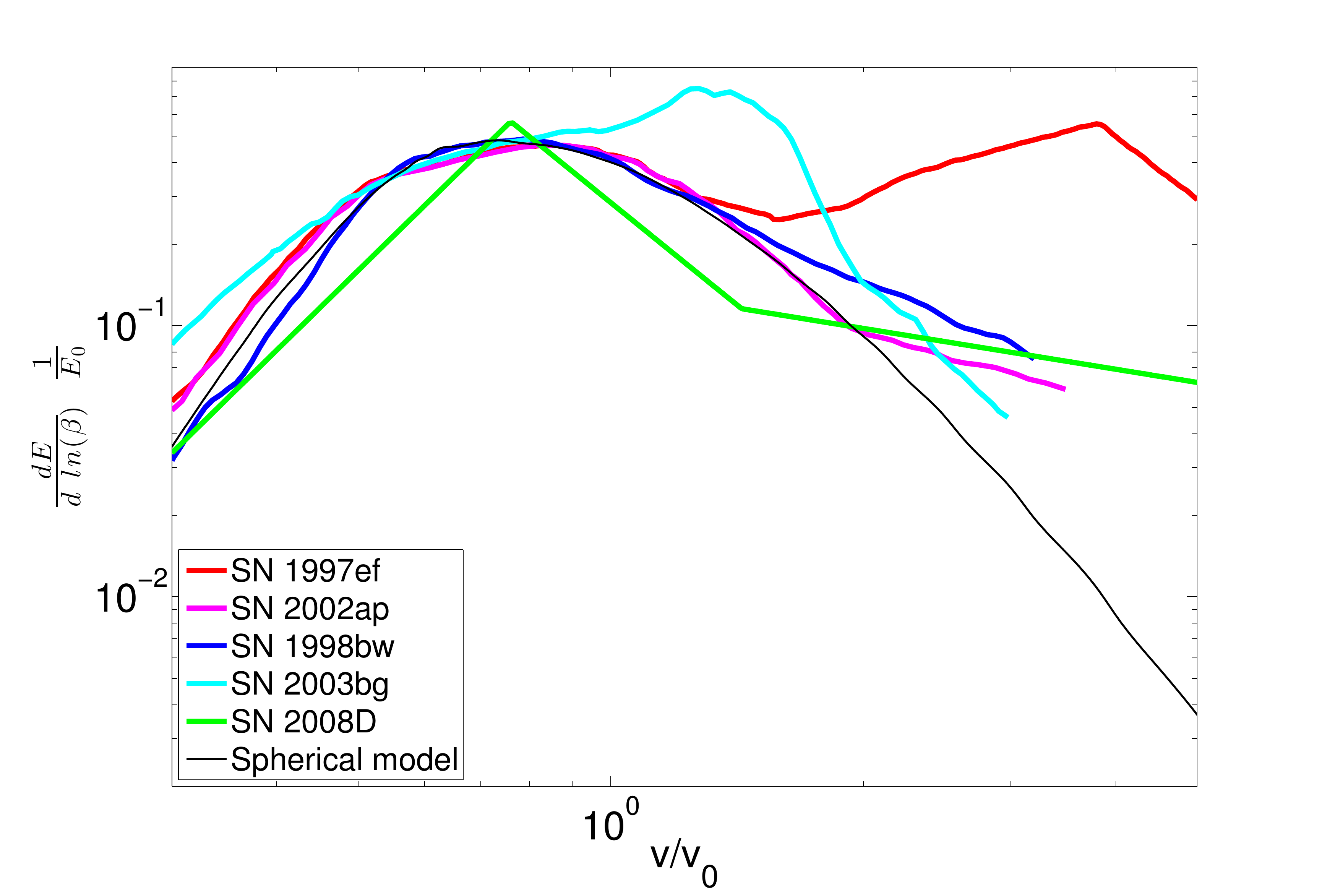}
\caption{The energy distribution of a function of the velocity for SNe in our sample. This plot does not include SN 2016jca which has a similar distribution to SN 1998bw.  All distributions are normalized ({each by different values of $v_0$ and $E_0$}) so the peaks of the distributions of the bulk of the ejecta coincide. The {\it thin black} line shows the distribution obtained from a numerical simulation (using the code PLUTO \cite{mignone2007}) of a spherical explosion  of a progenitor with a standard  pre-expolsion density profile near the stellar edge, $\rho (r) \propto (R_*-r)^3$, where $R_*$ is the stellar radius. All SNe show an excess of material at high velocities which is not expected in the spherical model. Instead it is naturally explained by a powerful relativistic jet that deposits all its energy in a small amount of stellar mass falling within its cocoon opening angle.}
\label{fig:dEdv}%
\end{figure}

As a spherical shock wave generated at the center of the collapsing star propagates outwards it encounters a sharp density drop near the edge of the star. The shock then accelerates until it breaks out from the star. As the shock accelerates it loses causal contact with the energy reservoir behind it, depositing less and less energy, $E$, into progressively faster and faster material with velocity $v$. 
Regardless of the exact density profile near the stellar edge the acceleration of the shock results in a rapidly decreasing profile of $E(>v)$. For a typical envelop structure the fastest moving material satisfies $dE/d\log(v) \propto v^{-k}$ where $5 \le k \le 8$ \cite{matzner1999} (see Fig. \ref{fig:dEdv}).

\begin{figure}[!ht]
\includegraphics[width=1\textwidth]{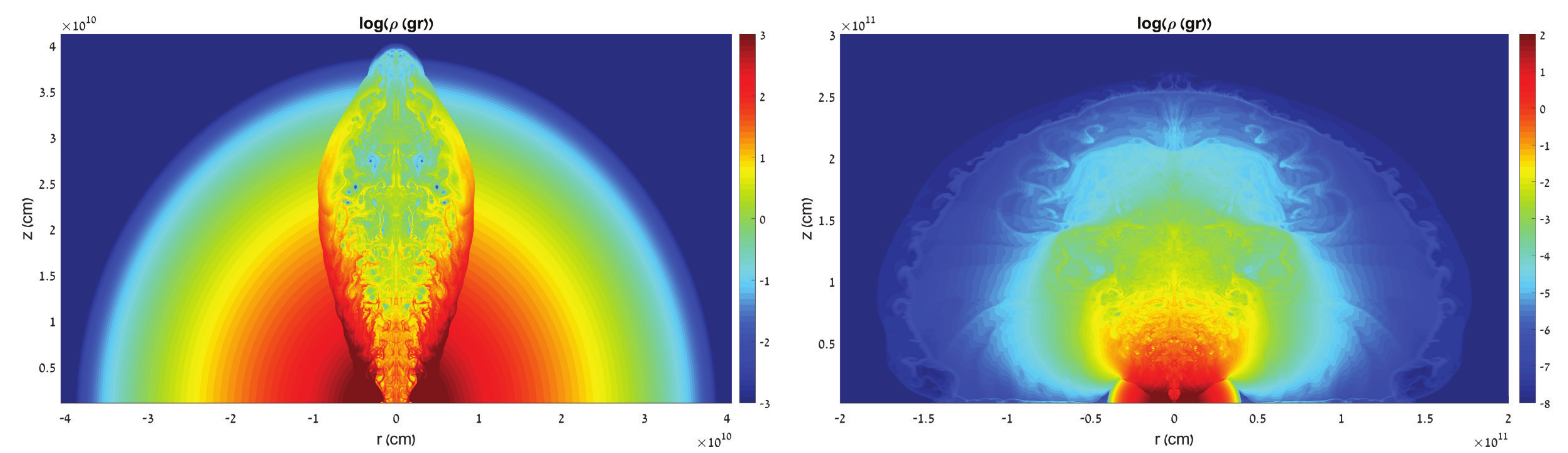}
\caption{Two snapshots from a relativistic hydrodynamic simulation of a choked relativistic jet done using the code PLUTO \cite{mignone2007} (from Gottlieb \& Nakar in preparation). The jet, with an opening angle of $8^{o}$, is choked when it is halfway through the stellar envelope. At that time the cocoon opening angle is  similar to the jet opening angle. At the time of breakout (right panel) the cocoon {half} opening angle is $\theta_c \approx 20^{o}$. After the breakout the cocoon material spills out of the star and spread in all directions (left panel). The simulation includes only a jet and it does not include the more spherical SN explosion.}
\label{fig:simulation}%
\end{figure}

As a relativistic jet carves its way through the stellar envelope  a  double shock (forward- reverse) structure forms at its head \cite{Matzner03, Lazzati05, B11}. The head propagates with a velocity much  slower than the jet itself. For typical jet-star parameters seen in GRBs this velocity is mildly relativistic \cite{ZhangHegerWossly}. The hot
head material spills sideways, forming a cocoon that engulfs the jet and collimates it. 
 As long as the jet propagates in the stellar envelope  it dissipates its energy  at the head. This energy flows into the cocoon. The jet continues to propagate as long as the engine driving it
operates. If it operates long enough the jet breaks out and powers a GRB. Otherwise the jet stalls
and all its energy, $E_j$,
is deposited into the cocoon.   At that time the cocoon contains the stellar mass within a cone with a {half}  opening angle $\theta_j$ \cite{B11}. 
The cocoon, that  is much hotter than the surrounding matter, expands and breaks out from the star.  If the jet has propagated a significant fraction of the stellar radius before it stalled 
the cocoon half opening angle at the time of breakout, $\theta_c$, is comparable to that of the original jet, $\theta_j$. Otherwise it could be much wider. 
As the cocoon's hot material breaks out from the star its  optical depth $\tau > c/v$ hence it  expands rapidly sideways and it engulfs the star (see Fig. \ref{fig:simulation}) reaching a  velocity of order $v_c\approx 0.1 c \sqrt{E_{j,51.5}  /  M_{10}\theta_{c,{10^o}}^2 }$, { where $E_j$ is the jet's total energy (that has been deposited in the cocoon) and $M$ is the stellar mass  \cite{NP17}. Here and elsewhere $Q_x$ denotes $Q/10^x$ in cgs  while $M_x$ is in units of solar mass.} The radiation escapes from the expanding cocoon material when it reaches $\tau \approx c/v$ at $ t_{obs} \approx  1.5 \kappa_{-1.3}^{1/2} M_{10}^{3/4}  \theta_{c,{10^o}}^{3/2} / E_{51.5}^{1/4}$ day,
where $\kappa$ is the opacity per unit mass. The luminosity at this time is 
$\approx  1.5 \times  10^{42} E_{51.5}   R_{11}  \theta_{c,{10^o}}^{4/3} / M_{10} \kappa_{-1.3}$ erg  and the  temperature is
$ \approx 12,000   E_{51.5}^{1/8}
 R_{11}^{1/4}/   \theta_{c,{10^o}}^{7/12}  M_{10}^{3/8} \kappa_{-1.3} {\rm ~K }$, {where $R$ is the progenitor radius.}
This rather uv/blue signal might be observed  if the SN is caught sufficiently early (but it might already be hidden by the rising  $^{56}$Ni decay driven emission).

While the direct emission signal is short-lived the cocoon's signature  on the velocity structure of the ejecta  can be observed for a longer period via absorption. During the first few days the absorption lines of the fast moving cocoon material are optically thick, thereby leaving their mark on the optical spectra. As a result, the density profile required to fit the early spectra of a SN with a significant jet activity is expected to show an additional very fast component (with $v \approx 0.1c$). {Namely, a flattening or a `bump' of the $E(v)$ (or equivalently $\rho(v)$) profile around this velocity, instead of the rapidly decreasing profile of a regular spherical explosion.} This signature can be seen only during the first few days since once the cocoon lines in the optical become optically thin this spectral signature disappears.
   
At high velocities the observed $E(v)$ profile is the sum of the rapidly decreasing regular SN energy profile  and the high velocity cocoon component.  The latter reflects the jet properties and it depends on the jet parameters: the energy and the opening angle as well as the depth at which the jet is choked. Clearly, a  less energetic, wider or a deeply choked jet will give rise to a less energetic and slower cocoon whose contribution will be weaker and at lower velocities and thus more difficult to separate from the bulk of the SN ejecta. 

\begin{table}\label{tbl:SN}
\begin{tabular}{|c|c|c|c|c|c|c| >{\centering}m{3cm}|c|}
    \hline
	SN & Type & $E_{tot}$ & $M_{ej}$ & $E_j$ & $M_c$  & $\theta_c$ &Comments& ref. \\
	 & & [$10^{51}$ erg] & $[M_\odot]$ & [$10^{51}$ erg]& $[M_\odot]$  & [deg] && \\
	\hline \hline
	1997ef & Ic-BL  &20 & 8 & 9 & 0.4 & $20^{o}$ & {\scriptsize No associated GRB}&\cite{mazzali2000}\\
	\hline
	1998bw & Ic-BL & 50 & 11 & $\gtrsim 2$  & - & - & {\scriptsize Associated with a low luminosity GRB 980425} & \cite{iwamoto1998}  \\
	\hline
	2002ap & Ic-BL & 4 & 2.5 & 0.3 & - & - & {\scriptsize No associated GRB. No outflow faster than 0.3c.}&\cite{mazzali2002}\\
	\hline
	2003bg & IIb &5 & 4.5 & 1 & 0.2 & $20^{o}$ & &\cite{mazzali2009}\\
	\hline
	2008D & Ib &6 & 7 & 1.4 & - &  - & {\scriptsize Associated with a faint  x-ray burst}&\cite{mazzali2008}\\
	\hline
	2016jca & Ic-BL & 50 & 10 & $\gtrsim 2$ & - & - & {\scriptsize Associated with a long GRB 161219b} & \cite{ashall2017}  \\
	\hline
\end{tabular}
\caption{Properties of the SNe in our sample. $E_{tot}$ and $M_{ej}$ are the  total SN energy and ejected mass. $E_j$ and $M_c$ are the energy and the mass excess of material moving at high velocities over the prediction of a spherical explosion (see Fig. \ref{fig:dEdv}) {and $\theta_c$ is the corresponding half opening angle of the cocoon upon breakout}. In all SNe, except for SNe 1998bw and 2016jca, the energy of the high velocity material is only weakly sensitive to the exact spherical model. In SNe 1997ef and 2003bg  the inferred mass depends only weakly on the spherical model. In these SNe $M_c$ provides an estimate of {$\theta_c$}, which puts, in turn,  an upper limit on the jet opening angle. In SNe 1998bw and 2016jca the energy excess at high velocities depends somewhat on the underlying spherical models. Moreover, if this excess is due to cocoon material then most of the cocoon energy may be at velocities where the energy profile is dominated by the bulk of the ejecta, and therefore these $E_j$ values are  rough lower limits on the jet energies in these two SNe.  Note that all values in the table have been  calculated assuming spherical symmetry. As the expanding material is not expected to be fully spherically symmetric,  this introduces an uncertainty of a factor of a few in these estimates. }
\end{table}

An excess of high velocity material ($\gtrsim 0.1c$) compared to the expectation from a spherical model is observed in all the hydrogen-stripped SNe with available early spectra that we examined (see Fig. \ref{fig:dEdv} and Table 1). The prominence of the excess differ from one SNe to another, and so is the confidence that the observations cannot be explained by a spherical explosion. In all SNe the excess of fast ejecta is most naturally explained as the cocoon material. The strongest jet signature is observed in SN 1997ef {(see Figs. 1 and 2 of ref. \cite{mazzali2000} for the early spectra and Fig. 8 for the density profile)}. At low velocities the energy profile fits the theoretical model of a spherical explosion of a typical progenitor very well. However  an unexpected,  well separated, component dominate the energy profile at $v >25,0000$ km/s. The energy and mass of the fast component, which in this SN can be estimated relatively well, measure the jet energy and puts an upper limit on the jet opening angle. A less pronounced, yet clear, flattening  of the energy profile is seen in SN 2002ap (at $v\approx 30,000$ km/s) and SN 2008D (at $v \approx 17,000$ km/s). The flat energy profile enables a rather robust estimate of the fast component energy, but its mass which is dominated by material with velocity near the flattening point, cannot be well separated from that of the bulk of the ejecta. SN 2003bg exhibit a `bump' in the energy profile at $15,000<v<30,000$ km/s. The bump is seen near the peak of the energy distribution and not as a separate component as in SN 1997ef and therefore its identification as jet activity is less secure. Yet, such energy profile is not expected in a spherical explosion of a conventional progenitor and jet activity provides a good explanation. Finally, SN 1998bw and 2016jca do not show a flattening of the energy profile at high velocities, but they also don't show the expected steepening.  At $v>30,000$ km/s the energy profiles fall much slower than what is expected in a regular spherical SN. Thus, while not demonstrating clearly a powerful jet signature these profiles  show an excess of fast moving material indicating a likely  jet activity. At least in the case of SN 2016jca we know that a jet exists as it is associated with a regular long GRB.

{Most interesting is the variety of SN types in which jet signature is detected. These cover almost all types of  CCSNe from progenitors that lost all, or most, of their hydrogen envelope. 
SNe 1997ef and 2002ap are broad-line Ic which are not associated with any type of high energy emission. In particular, SN 2002ap, that took place at a distance of about 8 Mpc was observed extensively, showing no signs of a relativistic outflow. Radio and X-rays observed emission several days after the explosion indicate that the velocity of the fastest moving material in this SNe is $\sim 70,000$ km/s \cite{bjornsson2004}. In addition, it shows broad lines only in its early spectra while the lines in the later spectra (near and after the peak) are relatively narrow, similar to those observed in regular type Ic SNe.  
A jet signature is seen also in the relatively regular type Ib SN 2008D. It shows broad-lines at early times (produced by cocoon material according to our interpretation) which disappear at later times. SN 2008D also shows an early optical component with a luminosity of  $\sim10^{42}$ erg/s and a temperature of  $\sim 10,000$ K \cite{modjaz2009}, that fits the expected cooling cocoon emission discussed above. If the excess of material at high velocities in SN 2003bg is also interpreted as a cocoon material, then jets are active also in SN that lost most, but not all, of their H envelope (type IIb). Finally, we detect a less pronounced, yet possible  jet signature also in broad-line Ic SNe that are associated with GRBs. SN 2016jca is associated with a regular long GRB, and therefor we know that a jet must be active in this SN. SN 1998bw is associated with a low-luminosity GRB 980425, where the gamma-rays are fainter by 3-4 orders of magnitude than in regular long GRBs. Here the observed gamma-rays are most likely a result of a mildly relativistic shock breakout through an extended envelope \cite{nakar2012,nakar2015} and if a jet is active then it is most likely choked. Interestingly, the jet  that we infer from the optical spectra carried $\gtrsim 2 \times 10^{51}$ erg while the gamma-ray emission that preceded  SN 1998bw  carried $\sim 10^{48}$ erg and its radio emission indicates a mildly relativistic  ejecta with $\Gamma \sim 3$ that carried $\sim 10^{49}$ erg \cite{kulkarni1998}.}

To conclude, the most natural
interpretation of the energetic fast moving component observed in the early spectra of the SNe in our sample, is that this is the cocoon's matter that broke out from the envelope. In one case (SN 2008D) we might have even seen the direct thermal emission of this hot cocoon material. 
{his interpretation implies the existence of powerful jets within these SNe. These jets  carry  a significant fraction of the explosion energy that are   similar to those observed in typical long GRB jets.} The sample presented here is small. However, the absorption lines of the cocoon's fast moving material become rapidly optically thin 
and can be detected only if  good spectra is taken very early on.  There are not
many observations of this kind. Our sample comprise most SNe with stripped (or nearly stripped) H envelope with early spectra that were analyzed to constrain the profile of the fast moving ejecta. This suggests that a significant fraction of the core-collapse SNe, and possibly all those that have lost most or all of their H envelope, harbor choked jets.  


It is interesting to note that our interpretation of the existence of jets SNe is supported by other circumstantial, though less conclusive,  evidence. Remarkably,  spectropolarimetry of the optical light of SN 2002ap suggest that its ejecta contain a nonisotropic fast component with energy and velocity similar to those we find here \cite{totani2003}. More generally, double peaked oxygen nebular lines observed in a large fraction of the type Ib/c SNe. This feature  imply a significant asphericity in the oxygen distribution of most, and possibly all, type Ib/c SNe. \cite{Mazzali2005,Maeda2008,Taubenberger2009}. 
Additional indirect supporting evidence is the appearance of Ni in outer regions (i.e., high velocity) of the SNe in our sample as a jet that emerges from the inner parts of the core would  bring freshly synthesized Ni to the outer regions \cite{mazzali2000,iwamoto1998,mazzali2002,mazzali2009,mazzali2008,ashall2017}. {Finally, the structure of CCSNe remnants also support jet activity during the SN explosion \cite{grichener2017}}.

While the jets that we infer from the early optical spectra don't contain enough energy to drive the SN 
explosion, {they may be the smoking gun of what actually drives the explosion}. 
 First, and most important, is that a fast rotating core is almost certainly required and magnetic fields are also likely to play a major role in the explosion. Second, the jet activity which seems to be present in most type I SNe suggests a relation between the explosion mechanism of regular type I CCSNe  and the extremely energetic SNe associated with GRBs. This puts into question the ability of the popular neutrino driven explosion to be the mechanism that drives these SNe, and suggests that any model of CCSNe explosion mechanism (at least of type Ib,c) should be ale to produce an extremely energetic quasi-spherical explosion accompanied by a narrow and energetic relativistic jet. 
 
 The fact that our sample does not include regular type IIp SNe does not imply that we can show that these do not harbor choked jets, since the massive H envelope in this type of SNe is expected to choke not only the jet but also the cocoon, thereby washing out any jet signature from the early spectra. 
 {The observation that jets are ubiquitous in SN explosions suggests that low-metallicity that is implied from the location of long GRBs is not an essential ingredient for the activity of a central engine. } 
 Finally, we note that while the current sample is small upcoming  transient searches (ZTF, GAIA, LSST and others)  will enable us to detect regularly early SNe spectra. Those  will reveal in the near future the fraction of SNe that harbor jets and will shed new light on SNe engines.

Before concluding we note that the shocks involved in these hidden jets may be the source of  high energy neutrinos observed by IceCube \cite{Icecube2017}. Unlike low-luminosity GRB, another proposed source of hidden jets  \cite{nakar2015,Senno2016}, where some of the shocks in the jets are expected to be collisionless, in regular SNe hidden jets all shocks are expected to be radiation dominated. Therefore, they are less likely to be efficient particle acceleration sites and thus strong neutrino sources. Nevertheless,  it is possible that a small fraction of the energy dissipated in radiation mediated shocks is channeled into high energy neutrinos.
If, as we find here, relativistic jets are common in SNe  then their high abundance reduces significantly the required energy output in high energy neutrinos per event and enable much less efficient sources.  Furthermore, as these sources are optically thick the Waxman-Bahcall bound does not apply to them.  

Interestingly the hidden jets can also be  detectable sources of gravitational radiation. The acceleration of a relativistic jet produces gravitational radiation \cite{Birnholtz2013} that peaks at sub Hz frequencies.  Gravitational waves from a long GRB jet at 500 Mpc are below the detection limit of advanced LIGO but are detectable by the proposed sub Hz detector DECIGO \cite{Decigo}. However, depending on the parameters of the jet and in particular on its initial Lorentz factor and the duration of the acceleration phase, a hidden jet in a nearby SN taking place at 10Mpc might be detectable by advanced LIGO. 

We thank O. Gottlieb for figure \ref{fig:simulation}. This research was supported by the I-Core center of excellence of the CHE-ISF. TP was partially supported by an advanced ERC grant TReX and by a grant from the Templeton foundation. EN was was partially supported by an ERC starting grant (GRB/SN) and an ISF grant (1277/13).  PM and EP acknowledge kind hospitality in Israel by the Weizmann Institute for Science in Rehovot and the Hebrew University of Jerusalem.

\end{document}